# Dueling QR Codes: The Hyding of Dr. Jeckyl

**David A. Noever and Forrest G. McKee**
**PeopleTec, Inc., Huntsville, AL**
david.noever@peopletec.com       forrest.mckee@peopletec.com

**ABSTRACT**

The paper presents a novel technique for encoding dual messages within standard Quick Response (QR) codes through precise half-pixel module splitting. This work challenges fundamental assumptions about deterministic decoding in the ISO/IEC 18004:2015 standard while maintaining complete compatibility with existing QR infrastructure. The proposed two-dimensional barcode-attack enables angle-dependent message selection while maintaining compatibility with unmodified QR readers and the 100 million US mobile users who use their phone's built-in scanners. Unlike previous approaches that rely on nested codes, watermarking, or error correction exploitation, our method achieves true one-to-many mapping by manipulating the physical sampling process built into the QR standard. By preserving critical function patterns while bifurcating data modules, we create automated codes that produce different but valid readings based on camera viewing angle. Experimental results demonstrate successful implementation across multiple use cases, including simple message text pairs, complex URLs (nsa.gov/nasa.gov), and security test patterns for malware and spam detectors (EICAR/GTUBE). Our technique achieves reliable dual-message decoding using standard QR readers at module scales of 9-11 pixels, with successful angle-dependent reading demonstrated across vertical, horizontal, and diagonal orientations. The method's success suggests potential applications beyond QR code phishing ("quishing") including two-factor authentication, anti-counterfeiting, and information density optimization. The half-pixel technique may offer future avenues for similar implementations in other 2D barcode formats such as Data Matrix and Aztec Code.

*Keywords:* QR codes, spatial multiplexing, angle-dependent encoding, digital cryptography, machine vision, error correction codes

*Introduction.* The adoption of QR codes has grown 16% annually [1], particularly accelerated by the COVID-19 pandemic. The two-dimensional barcode garners attention as a future replacement for manual passwords and as convenient shortcuts for faster mobile keyboard entries. Statista [2] estimates that mobile QR code users in the United States will exceed 100 million, or 45% of US shoppers in 2025. Global QR code payment users are projected to increase from $11.2 billion in 2022 to nearly $51 billion by 2032 [1]. This breadth extends far beyond simple web links and marketing materials into mobile payments and critical health, security and verification applications. The ISO/IEC 18004:2015 standard [3] for QR codes establishes a strict one-to-one relationship between code and content. The goal of the standard is to guarantee that any QR Code conforming to ISO/IEC 18004:2015 produces a deterministic decoding—meaning that regardless of the reader used, the decoding outcome is unambiguous and consistent [3]. This deterministic behavior is fundamental to achieving interoperability and reliability in applications that use QR Codes. Yet recent innovations [4-7] reveal an intriguing challenge to this fundamental principle.

The development of angle-dependent dual-message QR codes [5] demonstrates a novel one-to-many mapping [6] that operates within the constraints of standard QR code readers while breaking from the traditional one-to-one paradigm (Figure 1). This previous work [5-7] has not provided an algorithm or automated generator for dual-messaging and instead relied on manual blending of paired QR codes [7].

The fundamental challenge in creating dual-message or lenticular QR codes stems from a direct conflict with the ISO standard's core assumption [3] that each QR code maps to exactly one message. This one-to-one relationship is so deeply embedded in the standard that it shapes everything from the basic encoding to error correction mechanisms. The present technique highlights a method for industrializing one-to-many mappings that works with standard QR readers, effectively creating two valid QR codes that coexist in the same physical space. Because the camera angle determines the destination, the method can be compared to perspective distortion methods in art or holography such as *tabula scalata* images, turning pictures, tilt or wiggle stereoscopes, or more whimsically, image "winkies" as providing depth illusion for the 1960's pop art company, Swank [8].

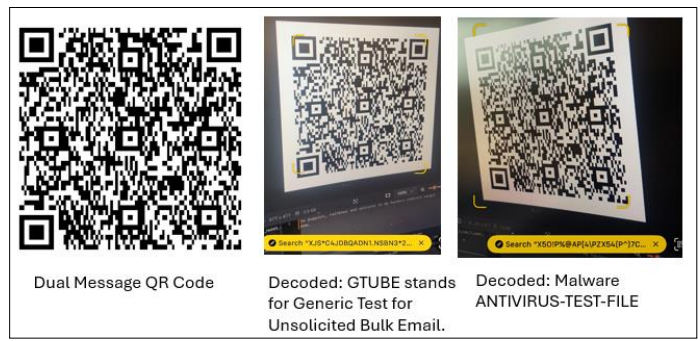

*Figure 1. Single QR Code Supports Dual Decoding to Security Test Patterns for Spam (L) and Virus Detectors (R)*

*Methods and Results.* This approach differs significantly from previous attempts [9-21] at embedding multiple messages in QR codes [16-20]. Earlier work focused on either nested codes, where one QR code contains another at a different zoom level [21] or embedding techniques that place logos or images within the code structure [9-10, 21]. Barron et al. [9,12] explored dual modulation techniques that require specific scanning conditions, while traditional security approaches focused on authentication through watermarking [10] or two-layer systems [20]. The physical principle of half-pixel splitting [5-6] represents a fundamental departure from these methods by creating a true one-to-many mapping that depends solely on viewing angle.

As shown in Figures 1-2, the innovation does not rely on 3D effects or nested codes [19-21], but rather on a deep understanding of how QR scanners sample pixels.

In brief, when the method generates one standard QR code, we preserve all the critical function patterns - the finder patterns, timing patterns, and alignment patterns - exactly as they appear in the first code. But for data modules, the software generator splits each module precisely in half, with the left half containing the value from the first QR code and the right half from the second. When this method scales the final image so each module is multiple pixels wide (typically 9-11 pixels), the *result creates a precise physical division that standard QR readers can interpret differently based on viewing angle.*

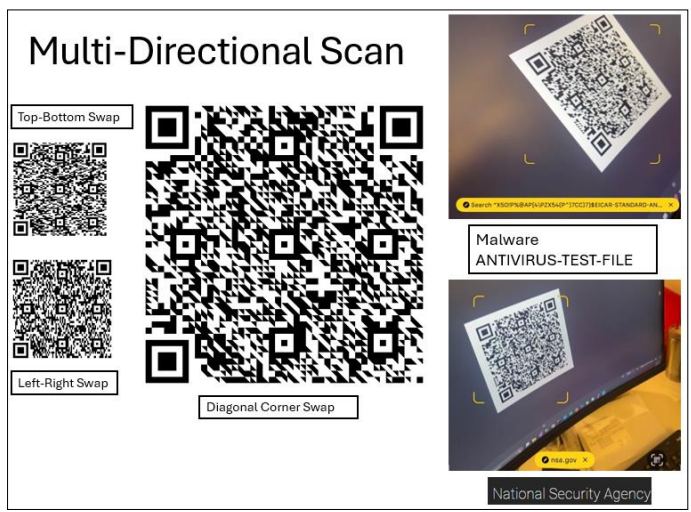

*Figure 2. Single QR Codes to Support Vertical, Horizontal and Diagonal Corner Scanning with Security Test Pattern for Virus Detection (Top) and Web URL nsa.gov (Bottom)*

As shown schematically in Figure 3, what makes this work is the fundamental nature of how QR readers operate [3]. The camera must first locate the code using finder patterns, establish a sampling grid based on timing patterns, and then sample near the center of each expected module position. The current half-pixel technique exploits this sampling behavior in a novel way. When viewed straight-on, the reader sees an ambiguous mix of both codes. But tilt the code slightly left, and the sampling points shift to fall on the left half of modules, reading the first QR code. Tilt right, and the sampling points shift to the right half, revealing the second code. In the present case, the QR code is both static and dual, encoding some probability of destination arrival based on the reader angle.

The dual-message QR code results demonstrate several key implementations and variations. Figure 1 highlights a test security case, where a scanner moving left to right or vice versa reveals a different version of the standard pair of test strings commonly used for penetration testing, namely the EICAR malware detection [22] and the GTUBE spam detection [23]. These test choices underscore the traditional QR destination risks called QR code phishing (or "quishing" in email spam as GTUBE) or the short link redirection from the user-initiated web page loading to a malicious download (or virus as EICAR).

Figure 2 shows a generalization of the perspective distortion. The images show the same half-pixel merging of the QR data modules works not just from side angles (Figure 1), but also from top to bottom and corner-to-corner diagonal scans. For concreteness, the diagonal QR code in Figure 2 illustrates the shift between the malware drop (e.g. EICAR string) and the link to the National Security Agency website (nsa.gov). These variations explore different geometric approaches to data module splitting, with implications for scanner compatibility and reading reliability.

This differs fundamentally from the standard QR approach where each module must unambiguously represent either black or white. Instead of hiding data within error correction capacity like traditional modifications, this algorithm creates modules that are valid for both codes simultaneously through precise physical partitioning. The dual message is not a trick of encoding or error correction; rather it is a physical manipulation that creates two valid sampling paths for standard readers, effectively turning viewing angle into a message selector. The result is a true dual-message state where both messages are fully present with their own error correction intact, challenging the validity of one-to-one mapping [3] while remaining compatible with existing QR infrastructure.

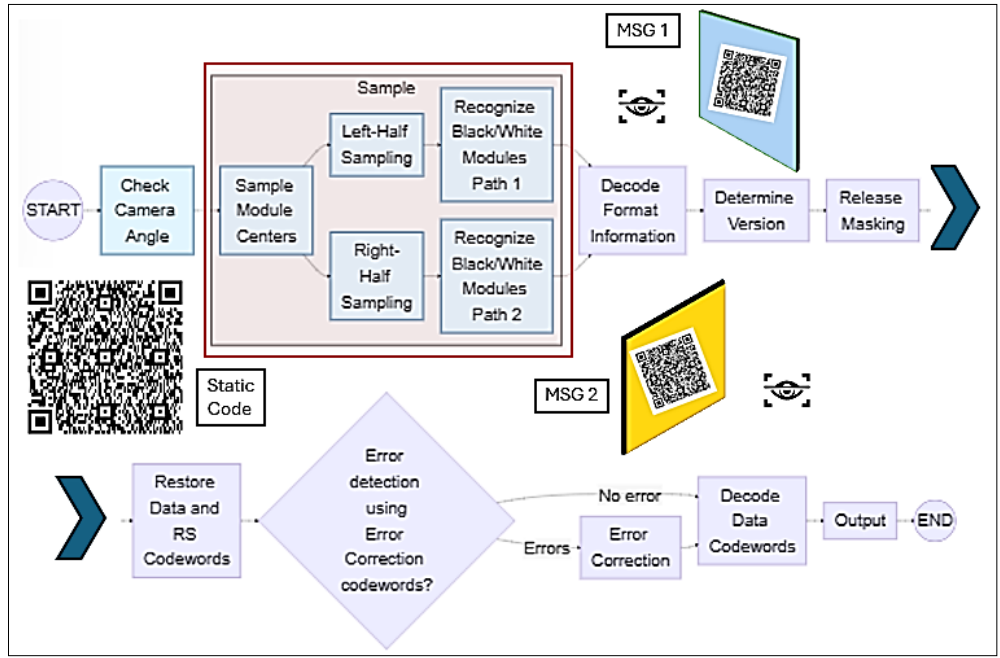

*Figure 3. The ISO Standard Deterministic Decoder vs the Modified Sampling (Red Tinted Box)*

*Discussion.* In Figure 3, the half-pixel technique primarily intersects with this decoding flow at the very first step: "Recognize Black/White Modules". By precisely splitting each data module, the algorithm creates a situation where the scanner's sampling points will consistently land on one half of the module depending on viewing angle as inclined left-right, top-bottom, or corner-to-corner. This means that before the decoder even reaches "Decode Format Information" or "Determine Version," it's already working with a complete, valid set of black/white module readings – they are just different readings depending on the viewing angle.

The error correction stage in this flow remains untouched and functional because each viewing angle produces a complete, valid QR code that can pass through error detection and correction normally. This is what makes the technique so robust against ISO standard defenses. The attack works by manipulating the physical sampling process rather than trying to exploit or modify the logical decoding steps that follow.

The extension of QR codes to dual-message capability [1,3] presents both opportunities and challenges for many high-trust applications. While it enables new security features like angle-dependent verification, it also requires careful consideration of potential exploit scenarios. In healthcare, QR codes have become integral to patient safety protocols. Studies show that QR-coded patient identification wristbands reduce medication errors by up to 57% compared to traditional methods in the quarter of US hospitals that use them [24]. Since its launch November 27, 2024, the Drug Supply Chain Security Act (DSCSA) mandates a comprehensive system of tracking, tracing, and verifying pharmaceutical products [25] based on labeling lot numbers, expiration dates and anti-counterfeit compliance.

The financial sector has seen particularly robust adoption, especially in Asia where over 95% of mobile phone users rely on QR codes as their primary payment methods across leading platforms like Alipay, Apple Pay and WeChat Pay. China alone had nearly 10 billion mobile devices that used QR codes for payments in 2022 [26].

The physical implementation of QR codes at architectural scale presents intriguing possibilities for our angle-dependent encoding technique. While Facebook's headquarters famously features a 42-foot QR code visible from satellite imagery [27], the present research suggests the possibility of creating massive dual-message codes that would resolve to different URLs depending on the satellite's orbital position and imaging angle (Figure 4). This target instance builds on existing examples of large-scale QR codes, such as agricultural crop patterns and rooftop advertisements, but adds a dynamic element through precise geometric design. This application would effectively transform static architectural QR codes from simple identifiers into dynamic, position-dependent information portals, creating what might be described as the first "orbital-position-dependent" digital markers.

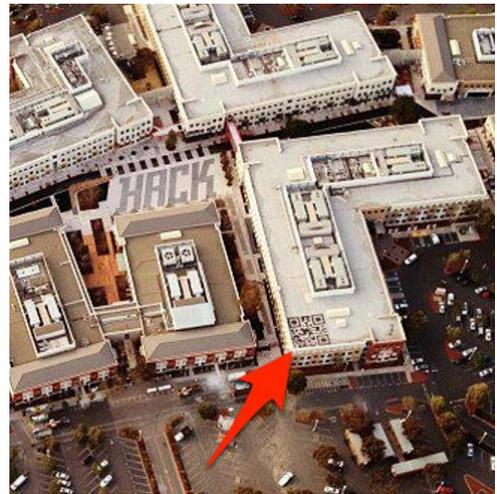

Figure 4. Satellite visible QR Codes on Facebook Building in 2012 Suggest Modified Destinations Depending on Camera Inclination

*Summary and Future Work.* The successful implementation of dual-message QR codes through half-pixel splitting opens fascinating possibilities for extending this technique to other barcode symbology (Figure 5). By challenging the one-to-one paradigm of 2D barcodes, similar approaches could work across various barcode types, though with differing levels of complexity and likelihood of success. The dominant consumer barcodes are linear and 2D QR codes. Data Matrix emerges as the most promising immediate candidate for adaptation. Its regular matrix structure and L-shaped finder pattern closely parallel QR codes, making it a candidate next step. The built-in error correction and square modules would likely respond well to the same half-pixel splitting technique, requiring only minor modifications to account for its different finder pattern structure.

Aztec Code presents an intriguing opportunity due to its unique central bullseye finder pattern and lack of quiet zone requirement. The concentric data rings could enable novel radial splitting patterns, potentially creating new ways to encode multiple messages. While the central finder pattern would need careful preservation, the high error correction capability of Aztec codes could provide robust tolerance for the splitting technique.

PDF417 represents a more challenging but still feasible target. Its stacked linear structure differs fundamentally from matrix codes, but this might enable new approaches to message splitting. The presence of start/stop patterns and row indicators suggests that vertical splitting between rows could offer a novel implementation path, though maintaining scanner compatibility would require careful consideration.

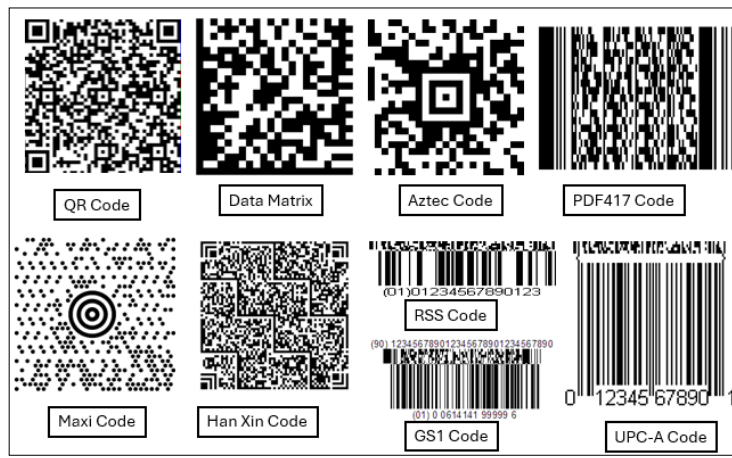

MaxiCode stands out as perhaps the most challenging yet intellectually interesting candidate. Its hexagonal module structure and bull's-eye finder pattern would require significant geometric innovation in the splitting technique. However, this very complexity might enable new forms of message encoding not possible with square-module codes.

The Han Xin Code, popular in Asia, shares many structural similarities with (diagonal)

Figure 5. Bar Code Symbology for Future Half-Pixel Tests

QR codes and could likely adopt similar half-pixel splitting techniques. Its high data density and strong error correction capabilities make it particularly suitable for dual-message implementation, potentially opening new applications in markets where this symbology is prevalent.

Composite symbologies like GS1 Composite and RSS Composite present special challenges but also unique opportunities. Their inherent layered structure might enable novel approaches to multiple message encoding, though the complexity of maintaining synchronization between components would require sophisticated solutions.

This work suggests a multi-faceted future research direction in barcode technology, potentially modifying how we think about international standard identifiers and information density in 2D codes. The implications extend beyond mere technical demonstration and algorithm validation, offering new possibilities in authentication systems, security watermarking, and anti-counterfeiting measures. As barcode technology continues to evolve, these multi-message techniques could become standard features rather than experimental innovations.

The primary technical constraint of angle-dependent QR codes centers on scanning inconsistency. Direct perpendicular scans produce unreliable results as sampling points fall on module boundaries, creating ambiguous readings that can resolve to either message or fail to decode entirely. In theory, user education and QR code placement could mitigate a confusing camera perspective on machine-read printed materials. Given that most QR codes are fixed mounted in commercial settings, a malicious actor could force a scan by inclining a stand at present angles such as 20 degrees to the vertical or horizontal depending on the target and goals. User education for QR codes also naturally support visual indicators for head-on views or directional markers to encourage a uniform perspective. In use cases such as mobile-to-mobile scans of identification, few of these methods would constrain the practical viewing angle and thus could spawn a random or chaotic destination. Spontaneous, any-angle scanning makes standard QR codes convenient for transferring information quickly and presumably consistently.

In conclusion, the fundamental architecture of QR codes, as specified in the ISO/IEC 18004:2015 standard, is built around a deterministic encoding and decoding process. Every aspect of the specification - from the structure of finder patterns to the arrangement of data modules and error correction coding - presumes a singular mapping between physical pattern and decoded content. This foundational assumption of one-to-one correspondence between code and content has shaped both the development of QR technology and its widespread adoption across industries. This research, however, demonstrates that this presumed constraint, while deeply embedded in both the standard and implementations, can be generally circumvented while maintaining compatibility with standard QR code readers. This challenges not just implementation assumptions but core principles about how machine-readable codes must function.

## ACKNOWLEDGEMENTS

The authors thank the PeopleTec Technical Fellows program for research support.

## REFERENCES

[1] Avinash, D. (2023), Global QR Codes Payment Market Size to Surpass USD 51.58 Billion Growth by 2032 | Surge in Adoption of Contactless Payments, https://www.fintechfutures.com/techwire/global-qr-codes-payment-market-size-to-surpass-usd-51-58-billion-growth-by-2032-surge-in-adoption-of-contactless-payments/

[2] Statista (2023), U.S. smartphone users scanning QR codes 2020-2025 https://www.statista.com/statistics/1297768/us-smartphone-users-qr-scanner/

[3] ISO/IEC 18004:2015 (2015) Information technology-Automatic identification and data capture techniques-QR Code bar code symbology specification, https://github.com/yansikeim/QR-Code/blob/master/ISO%20IEC%2018004%202015%20Standard.pdf

[4] Chou, K. C., & Wang, R. Z. (2024). Dual-Message QR Codes. *Sensors*, *24*(10), 3055.

[5] Williams, E. (2025) This QR Code Leads to Two Websites, But How?, Hackaday, https://hackaday.com/2025/01/23/this-qr-code-leads-to-two-websites-but-how/

[6] Dupont, G. (2025), Mastodon post, https://mstdn.social/@isziaui/113874436953157913

[7] Benchoff, B. (2011), Hacking QR Codes for Fun and Profit, Hackaday, https://hackaday.com/2011/08/09/hacking-qr-codes-for-fun-and-profit/

[8] Kim, H., Kim, M., & Lee, Y. (2021). Ambiguus Tiles: Origami Rhombic Pyramid Tiles for Creating Dual-View Tile Mosaics. *Leonardo*, *54*(2), 206-207.

[9] Barron, I., Yeh, H. J., Dinesh, K., & Sharma, G. (2020). Dual modulated QR codes for proximal privacy and security. *IEEE Transactions on Image Processing*, *30*, 657-669.

[10] Hsu, F. H., Wu, M. H., & Wang, S. J. (2012, September). Dual-watermarking by QR-code Applications in Image Processing. In *2012 9th International Conference on Ubiquitous Intelligence and Computing and 9th International Conference on Autonomic and Trusted Computing* (pp. 638-643). IEEE.
[11] Kammason, C., Wanna, Y., Wiratchawa, K., & Intharah, T. (2022, November). Dual Image QR Codes: The Best of Both Worlds. In *2022 International Conference on Digital Image Computing: Techniques and Applications (DICTA)* (pp. 1-8). IEEE.
[12] Barron, I. R., & Sharma, G. (2023). Optimized Modulation and Coding for Dual Modulated QR Codes. *IEEE Transactions on Image Processing*, *32*, 2800-2810.
[13] Escobedo, P., Ramos-Lorente, C. E., Ejaz, A., Erenas, M. M., Martínez-Olmos, A., Carvajal, M. A., ... & Palma, A. J. (2023). QRsens: Dual-purpose Quick Response code with built-in colorimetric sensors. *Sensors and Actuators B: Chemical*, *376*, 133001.
[14] Harini, N., & Padmanabhan, T. R. (2013). 2CAuth: A new two factor authentication scheme using QR-code. *International Journal of Engineering and Technology*, *5*(2), 1087-1094.
[15] Song, C., Li, Z., Xu, W., Zhou, C., Jin, Z., & Ren, K. (2018). My smartphone recognizes genuine QR codes! practical unclonable QR code via 3D printing. *Proceedings of the ACM on Interactive, Mobile, Wearable and Ubiquitous Technologies*, *2*(2), 1-20.
[16] Bui, T. V., Vu, N. K., Nguyen, T. T., Echizen, I., & Nguyen, T. D. (2014, August). Robust message hiding for QR code. In *2014 Tenth International Conference on Intelligent Information Hiding and Multimedia Signal Processing* (pp. 520-523). IEEE.
[17] Lin, P. Y., & Chen, Y. H. (2017). High payload secret hiding technology for QR codes. *EURASIP Journal on Image and Video Processing*, *2017*, 1-8.
[18] Huang, P. C., Li, Y. H., Chang, C. C., & Liu, Y. (2018). Efficient scheme for secret hiding in QR code by improving exploiting modification direction. *KSII Transactions on Internet and Information Systems (TIIS)*, *12*(5), 2348-2365.
[19] Tkachenko, I., Puech, W., Destruel, C., Strauss, O., Gaudin, J. M., & Guichard, C. (2015). Two-level QR code for private message sharing and document authentication. *IEEE Transactions on Information Forensics and Security*, *11*(3), 571-583.
[20] Yuan, T., Wang, Y., Xu, K., Martin, R. R., & Hu, S. M. (2019). Two-layer QR codes. *IEEE Transactions on Image Processing*, *28*(9), 4413-4428.
[21] Chou, G. J., & Wang, R. Z. (2020). The nested QR code. *IEEE Signal Processing Letters*, *27*, 1230-1234.
[22] Dunham, K. (2004). EICAR Test File Security Considerations. *Inf. Secur. J. A Glob. Perspect.*, *12*(6), 7-11.
[23] Both, D. (2023). Combating Spam. In *Using and Administering Linux: Volume 3: Zero to SysAdmin: Network Services* (pp. 319-348). Berkeley, CA: Apress.
[24] Khammarnia, M., Kassani, A., & Eslahi, M. (2015). The efficacy of patients' wristband bar-code on prevention of medical errors. *Applied clinical informatics*, *6*(04), 716-727.
[25] Suarez, V (2024), The National Security Need to Optimize the Drug Supply Chain Security Act (DSCSA), https://councilonstrategicrisks.org/2024/11/22/the-national-security-need-to-optimize-the-drug-supply-chain-security-act-dscsa/
[26] Statista, (2025), Top 10 countries worldwide based on number of mobile devices with QR codes 2022, https://www.statista.com/statistics/1388333/mobile-devices-that-accept-qr-code-payments-by-country/
[27] Constine, J (2012), Facebook Decorates Its Roof With A 42-Foot Wide QR Code, https://techcrunch.com/2012/03/25/facebook-rooftop-qr-code/

**Supplemental Materials: Pseudo-code Algorithm for Dual Message QR Codes**

```
ALGORITHM
  CreateDualQRCode(message1, message2, scale=9):
  // Initial QR Generation
  QR1 = GenerateQRCode(message1, version=7, error_correction=HIGH)
  QR2 = GenerateQRCode(message2, version=7, error_correction=HIGH)

  // Validate inputs
  AssertSameSize(QR1, QR2)
  AssertValidScale(scale)  // Must be even for clean splits

  // Create output canvas
  size = GetSize(QR1)
  outputSize = size * scale
  output = CreateBlankImage(outputSize, outputSize)

  // Process each module
  FOR y = 0 TO size-1:
      FOR x = 0 TO size-1:
          outputX = x * scale
          outputY = y * scale

          IF IsStructuralElement(x, y, size):
              // Preserve structural elements from QR1
              moduleValue = GetModuleValue(QR1, x, y)
              FillModule(output, outputX, outputY, scale, moduleValue)
          ELSE:
              // Split data modules
              module = CreateBlankModule(scale, scale)
              value1 = GetModuleValue(QR1, x, y)
              value2 = GetModuleValue(QR2, x, y)

              // Fill left half from QR1
              FillHalfModule(module, LEFT_HALF, value1)
              // Fill right half from QR2
              FillHalfModule(module, RIGHT_HALF, value2)

              PlaceModule(output, outputX, outputY, module)

  RETURN output

FUNCTION IsStructuralElement(x, y, size):
    // Check if position is part of:
    // 1. Finder patterns (including quiet zone)
    IF IsInFinderPattern(x, y, size): RETURN true

    // 2. Timing patterns
    IF x == 6 OR y == 6: RETURN true

    // 3. Format information areas
    IF IsInFormatArea(x, y, size): RETURN true

    // 4. Version information areas (v7)
    IF IsInVersionArea(x, y, size): RETURN true

    // 5. Alignment pattern (v7)
    IF IsInAlignmentPattern(x, y): RETURN true

    RETURN false
```

```
FUNCTION IsInFinderPattern(x, y, size):
    finderLocations = [(0,0), (0,size-7), (size-7,0)]
    FOR each (fx,fy) in finderLocations:
        IF x in [fx-4 to fx+11] AND y in [fy-4 to fy+11]:
            RETURN true
    RETURN false
```